\documentstyle[aps,preprint]{revtex}

\begin{document}
\draft
\title{ bosonization theory of fermions interacting {\em via}
a confining potential
}
\author{Tai-Kai Ng}
\address{
Department of Physics,
Hong Kong University of Science and Technology,\\
Clear Water Bay Road,
Kowloon, Hong Kong
}
\date{ \today }
\maketitle
\begin{abstract}
   We study in this paper the properties of a gas of fermions interacting
{\em via} a scalar potential $v(q)=4\pi{e}^2/q^2$ for $q<\Lambda<<k_F$
at dimensions larger than one, where $\Lambda$ is a high momentum cutoff
and $k_F$ is the fermi wave vector. In particular, we shall consider
the $e^2\rightarrow\infty$ limit where the potential becomes
confining. Within a bosonization approximation, effective
Hamiltonians describing the low energy physics of
the system are constructed, where we show that the system can be 
described as a fermi liquid formed by chargeless quasi-particles which 
has vanishing wavefunction overlap with the bare fermions in the system.

\end{abstract}


\narrowtext

\section{introduction}
    In the last few years, there has been enormous interests in the
study of $U(1)$ gauge theories of fermionic systems in dimensions
higher than one, as a result of appearance of effective gauge
theories in the $t-J$ type models\cite{t1} and 
also in the studies of the $\nu={1\over2}$ Fractional Quantum
Hall state\cite{t2,hal}. In particular, the behaviour of the systems in the
confinement regime where charge excitations are "confined" are of
special interests. In this paper we shall apply a bosonization 
procedure to study a system of charge $e$ spinless fermions
interacting with scalar gauge field $\phi$ in dimensions
higher than one, with effective long distance gauge field action $L_{\phi}=(\nabla\phi)^2$. The problem is equivalent to fermions 
interacting {\em via} scalar potential $v(q)=4\pi{e}^2/q^2$ for $q<\Lambda$,
where $\Lambda<<k_F$ is a high momentum cutoff, $k_F$ is the fermi
wave vector. In particular we are interested
at the $e^2\rightarrow\infty$ limit where the potential
becomes confining. In this limit any non-uniform charge density
fluctuations $<\rho(\vec{q})>\neq0$ with any wave vector $q<\Lambda$ cost
infinite electric field energy. As a result any physical state 
$|\psi>$ which survives in this limit satisfies the constraint 
$\rho(\vec{q})|\psi>=0$, where $\rho(\vec{q})$ is the charge density
operator. 

  In this paper we shall show that within a bosonization 
approximation the projection to the physical states
$|\psi>$ where density fluctuations are forbidden can be 
carried out in the $e^2\rightarrow\infty$ limit, the 
bosonization approximation being a generalization of usual
bosonization method in one dimension to dimensions higher than
one\cite{b0,b1,b2,b3}. Within the approximation 
effective low energy Hamiltonians
which describes the particle-hole as well as
single-particle excitations in the system can be constructed.
In particular, the single particle excitations are described by 
operators which commutes with charge operator $\rho(\vec{q})$, and
has zero-overlap with the bare fermion operators in the system. 
The ground state of the system can be described as a fermi liquid
of these new single particle operators and corresponding to a
kind of 'marginal' fermi liquid in the original fermion description.
The organization of our paper is as follows: in section two
we shall outline our mathematical formulation of the bosonization
procedure where the nature of the bosonization approximation
will be explained. In section III we shall study in
detail the eigenstates and eigenvalues of the particle-hole
excitation spectrum in the bosonization approximation for arbitrary 
values of coupling constant $e^2$. In particular we shall show
how the projection to the physical Hilbert space where charged
excitations are forbidden is archieved in the 
$e^2\rightarrow\infty$ limit. In section IV we shall
consider the single particle excitations where 'physical'
single particle operators which commute with density operators
are constructed and the equations of motion for these single-particle 
operators are derived. We shall show that the system can be described as
a marginal fermi liquid of the original fermions. Our results
will be summarized in section V where some further comments will be given. 
 
\section{mathematical formulation}
   We consider a gas of spinless fermions interacting {\em via}
scalar potential $v(q)=4\pi{e}^2/q^2$ in momentum space. The Hamiltonian 
of the system is, 
\begin{equation}
\label{ham}
H=\sum_{\vec{k}}\epsilon_{\vec{k}}f^+_{\vec{k}}f_{\vec{k}}+{1\over2L^d}
\sum_{\vec{q}\neq0,q<\Lambda}v(q)\rho(\vec{q})\rho(-\vec{q}),
\end{equation}
where $\epsilon_{\vec{k}}=(\vec{k})^2/2m$ and $f(f^+)_{\vec{k}}$'s
are fermion annihilation(creation) operators. We have set $\hbar=1$ to
simplify notation. $\rho(\vec{q})=\sum_{\vec{k}}f^+_{\vec{k}+\vec{q}/2}
f_{\vec{k}-\vec{q}/2}$ is the density operator for the fermions. $L^d$
is the volume of the system. $\Lambda<<k_F$ is a high momentum cutoff. 
Notice that the Hartree ($\vec{q}=0$) interaction energy does not 
appear in the Hamiltonian as in usual Coulomb gas problem. The 
difference between our model and usual Coulomb gas problem has to be 
emphasized here. In usual Coulomb gas problem the
high momentum cutoff $\Lambda$ is taken to be infinity, or satisfies
$\Lambda>>k_F$. In this limit a Wigner crystal is expected to be formed
(in 3D) when $e^2$ becomes large because the potential energy term becomes
dominating at length scale $\sim{k}_F^{-1}$. In our problem where 
$\Lambda<<k_F$, the interaction is effective only at length scale $>>$
inter-particle spacing and formation of Wigner crystal is not possible 
since the fermions are essentially non-interacting at distance scale
of inter-particle spacing. Instead the system should remain in a liquid 
state and a new treatment to the problem is 
necessary. We note that bosonization method is a natural tool to attack 
the problem in this limit\cite{b2,b3,b4}. The bosonization procedure 
can be formulate most easily by introducing the Wigner function operators 
$\rho_{\vec{k}}(\vec{q})=f^+_{\vec{k}+\vec{q}/2}f_{\vec{k}-\vec{q}/2}$.
We shall work in the path-integral formulation where the Wigner
operators are introduced in the action of the system at imaginary
time through Langrange multiplier fields,
\begin{eqnarray}
\label{lag}
S & = & \int^{\beta}_0d\tau\left[\sum_{\vec{k}}f^+_{\vec{k}}(\tau)
({\partial\over\partial\tau}+\epsilon_{\vec{k}}-\mu)f_{\vec{k}}(\tau)-
\sum_{\vec{k},\vec{q}\neq0}i\lambda_{\vec{k}}(\vec{q},\tau)(\rho_{\vec{k}}
(\vec{q},\tau)-f^+_{\vec{k}+\vec{q}/2}(\tau)f_{\vec{k}-\vec{q}/2}(\tau)
)\right.  \\   \nonumber
& & +\left.{1\over2L^d}\sum_{\vec{q}\neq0,q<\Lambda,\vec{k},\vec{k}'}v(q)
\rho_{\vec{k}}(\vec{q},\tau)\rho_{\vec{k}'}(-\vec{q},\tau)\right],
\end{eqnarray}
where $\mu$ is the chemical potential.
$\lambda_{\vec{k}}(\vec{q})$ are Lagranger multiplier fields introduced
to enforce the constraint that the Wigner operators are given by
$\rho_{\vec{k}}(\vec{q})=f^+_{\vec{k}+\vec{q}/2}f_{\vec{k}-\vec{q}/2}$.
In particular, the original Hamiltonian \ (\ref{ham}) is recovered once
the $\lambda_{\vec{k}}(\vec{q})$ field is integrated out. 

  Integrating out first the fermion fields $f(f^+)$'s we obtain
an action in terms of $\rho_{\vec{k}}(\vec{q})$ and
$\lambda_{\vec{k}}(\vec{q})$ fields,
\begin{equation}
\label{lan2}
S={1\over\beta}F_0-Trln\left[\hat{1}-\hat{G}_o\hat{\lambda}\right]
-\sum_{\vec{k},\vec{q}\neq0,q<\Lambda}i\lambda_{\vec{k}}
(\vec{q})\rho_{\vec{k}}(\vec{q})+{1\over2L^d}
\sum_{\vec{q}\neq0,q<\Lambda,\vec{k},\vec{k}'}v(q)
\rho_{\vec{k}}(\vec{q})\rho_{\vec{k}'}(-\vec{q}),
\end{equation}
where $F_0$ is the free energy for an non-interacting Fermi gas.
$\hat{G}_0$ and $\hat{\lambda}$ are infinite matrices in
wave vector and frequency space, with matrix elements given by
\begin{equation}
\label{green}
\left[\hat{G}_0\right]_{k,k'}=\delta_{k,k'}g_0(k)\; \;\;\;
g_0(k)={1\over{i}\omega_n-\xi_{\vec{k}}},
\end{equation}
and
\begin{equation}
\label{lambda}
\left[\hat{\lambda}\right]_{k,k'}={i\over\sqrt{\beta}}
\lambda_{{\vec{k}+\vec{k}'\over2}}(\vec{k}-\vec{k}',
i\omega_n-i\omega_{n'})={i\over\sqrt{\beta}}\lambda_{k+k'\over2}(k-k'),
\end{equation}
where $k=(\vec{k},i\omega_n)$ and $\xi_{\vec{k}}=\epsilon_{\vec{k}}-
\mu$. The $Trln\left[1-G_0\lambda\right]$ term can be expanded in
a power series of $i\lambda_{\vec{k}}(\vec{q})$ field,
\[
Trln\left[\hat{1}-\hat{G}_o\hat{\lambda}\right]=-Tr\left[\hat{G}_0
\hat{\lambda}\right]-{1\over2}Tr\left[\hat{G}_0\hat{\lambda}\right]^2
-{1\over3}Tr\left[\hat{G}_0\hat{\lambda}\right]^3+O(\hat{\lambda}^4).  \]
Keeping terms to second order in $\hat{\lambda}$ (Gaussian approximation), 
we obtain
\begin{equation}
\label{gauss}
Trln\left[\hat{1}-\hat{G}_0\hat{\lambda}\right]\sim{1\over2
\beta}\sum_{k,q}g_0(k+{q\over2})g_0(k-{q\over2})\lambda_{k}(q)
\lambda_k(-q).
\end{equation}
Notice that the first order term in $\hat{\lambda}$ gives the usual
Hartree self-energy and is excluded in the problem. The
$i\lambda_{\vec{k}}(\vec{q})$ fields in Action \ (\ref{lan2}) can
be integrated out in Gaussian approximation, resulting in an quadratic
action in terms of $\rho_{\vec{k}}(\vec{q})$ fields only. We obtain
\begin{equation}
\label{srho}
S_{\rho}={1\over2L^d}\sum_{\vec{k},\vec{k}',\vec{q},i\omega_n}
\left[-{1\over\chi_{0\vec{k}}(\vec{q},i\omega_n)}(L^d\delta_{\vec{k},
\vec{k}'})+v(q)\right]\rho_{\vec{k}}(\vec{q},i\omega_n)
\rho_{\vec{k}'}(-\vec{q},-i\omega_n),
\end{equation}
where
\begin{equation}
\label{x0}
\chi_{0\vec{k}}(\vec{q},i\omega_n)={1\over\beta}\sum_{i\Omega_n}
g_0(\vec{k}+\vec{q}/2,i\omega_n+i\Omega_n)g_0(\vec{k}-\vec{q}/2,i\Omega_n)
={n_{\vec{k}-\vec{q}/2}-n_{\vec{k}+\vec{q}/2}\over
i\omega_n-{\vec{k}.\vec{q}\over{m}}},
\end{equation}
$n_{\vec{k}}=\theta(-\xi_{\vec{k}})$ at zero temperature is the free
fermion occupation number. $S_{\rho}$ can be expressed in terms of 
canonical boson fields by introducing
\begin{mathletters}
\label{can}
\begin{equation}
\label{can1}
\rho_{\vec{k}}(\vec{q},i\omega_n)=\sqrt{|\Delta_{\vec{k}}(\vec{q})|}
\left(\theta(\Delta_{\vec{k}}(\vec{q}))a^+_{\vec{k}}(\vec{q},i\omega_n)
+\theta(-\Delta_{\vec{k}}(\vec{q}))a_{\vec{k}}(-\vec{q},-i\omega_n)\right),
\end{equation}
where $\Delta_{\vec{k}}(\vec{q})=n_{\vec{k}-\vec{q}/2}-n_{\vec{k}+
\vec{q}/2}$. Correspondingly, we also have
\begin{equation}
\label{can2}
\rho_{\vec{k}}(-\vec{q},-i\omega_n)=\sqrt{|\Delta_{\vec{k}}(\vec{q})|}
\left(\theta(\Delta_{\vec{k}}(\vec{q}))a_{\vec{k}}(\vec{q},i\omega_n)
+\theta(-\Delta_{\vec{k}}(\vec{q}))a^+_{\vec{k}}(-\vec{q},-i\omega_n)\right).
\end{equation}
\end{mathletters}
Putting eqs. \ (\ref{can}) back into $S_{\rho}$, we obtain after some
straightforward manipulations,
\begin{eqnarray}
\label{sa}
S_{\rho} & = & {1\over2}\sum_{\vec{k},\vec{q},i\omega_n}(-i\omega_n+
{|\vec{k}.\vec{q}|\over{m}})a^+_{\vec{k}}(\vec{q},i\omega_n)
a_{\vec{k}}(\vec{q},i\omega_n)+{1\over2L^d}\sum_{\vec{k},\vec{k}',
\vec{q},i\omega_n}v(q)\sqrt{|\Delta_{\vec{k}}(\vec{q})\Delta_{\vec{k}'}
(\vec{q})|}  \\  \nonumber
& & \theta(\Delta_{\vec{k}}(\vec{q}))\theta(\Delta_{\vec{k}'}(\vec{q}))
\times\left(a^+_{\vec{k}}(\vec{q},i\omega_n)a_{\vec{k}'}(\vec{q},
i\omega_n)+a^+_{\vec{k}}(\vec{q},i\omega_n)a^+_{-\vec{k}'}(-\vec{q},
-i\omega_n)\right.  \\ \nonumber
& & +\left.a_{-\vec{k}}(-\vec{q},-i\omega_n)a_{\vec{k}'}(\vec{q},
i\omega_n)+a_{-\vec{k}}(-\vec{q},-i\omega_n)a^+_{-\vec{k}'}
(-\vec{q},-i\omega_n)\right).
\end{eqnarray}
Notice that $S_{\rho}$ is an action for interacting bosons
described by boson fields $a(a^+)_{\vec{k}}(\vec{q})$ satisfying
usual boson commutation relations $[a_{\vec{k}}(\vec{q}),a^+_{\vec{k}'}
(\vec{q}')]=\delta_{\vec{k}\vec{k}'}\delta_{\vec{q}\vec{q}'}$ and
$[a_{\alpha},a_{\beta}]=[a^+_{\alpha},a^+_{\beta}]=0$ and
with kinetic energies $|\vec{k}.\vec{q}|/m$ and interaction term
of form $\sim{v}(q)(a^+a+a^+a^++aa+aa^+)$. In this form the dynamics
of the original fermion system is described completely in terms of
boson fields(bosonized). Notice
that we have so far restricted ourselves to the Gaussian
approximation. Higher order interaction terms between
bosons will appear in a cumulant expansion of the $\lambda_{\vec{k}}
(\vec{q})$ fields\cite{b4}. We shall assume in the following that
these high-order terms do not modify qualitatively the physics 
described by the Gaussian theory. Notice also that in the $\vec{q}
\rightarrow0$ limit, $\Delta_{\vec{k}}(\vec{q})\rightarrow-\delta
(\epsilon_{\vec{k}}-\mu)({\vec{k}.\vec{q}\over{m}})$ and the usual
"tomographic" bosonization procedure based on subdivision of Fermi 
surface into disjoint patches at small
$\vec{q}$ is recovered\cite{b1,b2,b3}. Our bosonization
procedure can be viewed as a generalization of the tomographic
bosonization method for small wave vector $\vec{q}$ to arbitrary 
values of $\vec{q}<\Lambda$.

  To understand the nature of bosonization theory and Gaussian 
approximation, we first evaluate the free energy associated with
$S_{\rho}$. Integrating out the $\rho_{\vec{k}}(\vec{q})$ or
$a(a^+)_{\vec{k}}(\vec{q})$ fields in $S_{\rho}$ and using the fact that
\[
Tr\left[v(q)\hat{\chi}_0(\vec{q},i\omega)\right]^n=
(v(q)\chi_0(\vec{q},i\omega))^n, \]
where $\left[\hat{\chi}_0(\vec{q},i\omega)\right]_{\vec{k},\vec{k}'}=
\chi_{0\vec{k}}(\vec{q},i\omega_n)$ and $\chi_0(\vec{q},i\omega_n)$
is the usual Lindhard function, we obtain
\begin{equation}
\label{frpa}
F_{\rho}={1\over2\beta}\sum_{\vec{k},\vec{q},i\omega_n}ln(
\chi_{0\vec{k}}(\vec{q},i\omega_n))+{1\over2\beta}\sum_{\vec{q},
i\omega_n}ln\left(1-v(q)\chi_0(\vec{q},i\omega_n)\right),
\end{equation}
where the first term in $F_{\rho}$ is coming from the kinetic energy
of the $a(a^+)$ bosons and the second term is the usual
random-phase approximation (RPA) correction to free energy 
for interacting fermions. The
presence of the RPA term in $F_{\rho}$ suggests that our present
Gaussian approximation is essentially the same as RPA theory for
interacting fermions, except that the extra 'bosonization' assumption
in our theory which gives rise to the extra kinetic term in 
$F_{\rho}$. The 'bosonized' action $S_{\rho}$ assumes that the
excitations in a fermion system can be fully represented by particle-hole
pairs which are treated as independent bosons in the Gaussian
apprioximation. Note that particle-hole pairs are not all
{\em independent} in a fermion system because of Pauli exclusion
principle. For example, two particle hole pairs 
$f^+_{\vec{k}}f_{\vec{p}}$ and $f^+_{\vec{k}}f_{\vec{p}'}$ are not
independent excitations because they both involve creation of fermions
in state $\vec{k}$. However, they are treated as independent bosons
here as long as $\vec{p}\neq\vec{p}'$. Notice, however that
in one dimension the situation becomes different when the fermion spectrum 
is linearized near the Fermi surface. In this limit the entire particle-hole
excitation spectrum can be represented rigorously by bosons\cite{b1} and the 
Gaussian approximation becomes "exact". At higher dimensions, it is also
believed that the bosonization approximation is good as long as the 
interaction cutoff satisfies $\Lambda<<k_F$\cite{b2,b3,b4}, and as long as
interaction with transverse gauge field is excluded\cite{gauge}.

   Despite these approximations in the Gaussian theory,
the bosonized form of the action has the
advantage that within the approximation the full excitation spectrum 
and the ground and excited states wavefunctions of the system
can be obtained easily. This allows us to study the properties of
the system in great detail, as we shall see
in the following.

\section{particle-hole excitation spectrum}
   The eigenstates and eigenvalue spectrum described by
the action $S_{\rho}$ can be obtained by diagonalizing
the bosonized action \ (\ref{sa}) using a generalized
Bogoliubov transformation. We introduce for each wave vector
$\vec{q}$ the Bogoliubov transformation\cite{b5}
\begin{eqnarray}
\label{bg}
a_{\vec{k}}(\vec{q}) & = & \sum_{\vec{k}'}\left[\alpha_{\vec{k}\vec{k}'}
\gamma_{\vec{k}'}(\vec{q})+\beta_{\vec{k}\vec{k}'}\gamma^+_{-\vec{k}'}
(-\vec{q})\right],  \\  \nonumber
a_{-\vec{k}}(-\vec{q}) & = & \sum_{\vec{k}'}\left[\alpha_{\vec{k}\vec{k}'}
\gamma_{-\vec{k}'}(-\vec{q})+\beta_{\vec{k}\vec{k}'}\gamma^+_{\vec{k}'}
(\vec{q})\right],
\end{eqnarray}
and with correspondingly,
\begin{eqnarray}
\label{bgi}
\gamma_{\vec{k}}(\vec{q}) & = & \sum_{\vec{k}'}\left[\alpha^*_{\vec{k}'
\vec{k}}a_{\vec{k}'}(\vec{q})-\beta_{\vec{k}'\vec{k}}a^+_{-\vec{k}'}
(-\vec{q})\right],  \\  \nonumber
\gamma_{-\vec{k}}(-\vec{q}) & = & \sum_{\vec{k}'}\left[\alpha^*_{\vec{k}'
\vec{k}}a_{-\vec{k}'}(-\vec{q})-\beta_{\vec{k}'\vec{k}}a^+_{\vec{k}'}
(\vec{q})\right],
\end{eqnarray}
where we assume that the $\gamma(\gamma^+)_{\vec{k}}(\vec{q})$ operators
diagonized the Hamiltonian, i.e.
\[
H_{\rho}=\sum_{\vec{k}}E_{\vec{k}}(\vec{q})\gamma^+_{\vec{k}}
(\vec{q})\gamma_{\vec{k}}(\vec{q})+E_G,   \]
where $E_{\vec{k}}(\vec{q})$ are the eigen-energies and $E_{G}$
is the ground-state
energy of the system. Notice that a collective mode may appear in the
system and is also included in the sum $\sum_{\vec{k}}$. The matrix
elements $\alpha$ and $\beta$ satisfies the orthonormality condition
\begin{eqnarray}
\label{on}
\sum_{\vec{k}"}\left[\alpha_{\vec{k}\vec{k}"}\alpha^*_{\vec{k}'\vec{k}"}
-\beta_{\vec{k}\vec{k}"}\beta^*_{\vec{k}'\vec{k}"}\right] & = &
\delta_{\vec{k}\vec{k}'},   \\  \nonumber
\sum_{\vec{k}"}\left[\alpha_{\vec{k}\vec{k}"}\beta_{\vec{k}'\vec{k}"}
-\beta_{\vec{k}\vec{k}"}\alpha_{\vec{k}'\vec{k}"}\right] & = & 0.
\end{eqnarray}
   Writing down the equation of motions for $a_{\vec{k}}(\vec{q})$'s
in terms of $\gamma(\gamma^+)_{\vec{k}}(\vec{q})$'s\cite{b5}, we obtain
the Bogoliubov equations
\begin{eqnarray}
\label{bgeq}
(E_{\vec{k}'}(\vec{q})-{|\vec{k}.\vec{q}|\over{m}})\alpha_{\vec{k}\vec{k}'}
& = & {v(q)\over{L}^d}\sum_{\vec{k}"}\theta(\Delta_{\vec{k}}(\vec{q}))
\theta(\Delta_{\vec{k}"}(\vec{q}))\sqrt{|\Delta_{\vec{k}}(\vec{q})
\Delta_{\vec{k}"}(\vec{q})|}(\alpha_{\vec{k}"\vec{k}'}+\beta^*_{\vec{k}"
\vec{k}'}),   \\ \nonumber
(E_{\vec{k}'}(\vec{q})+{|\vec{k}.\vec{q}|\over{m}})\beta_{\vec{k}\vec{k}'}
& = & -{v(q)\over{L}^d}\sum_{\vec{k}"}\theta(\Delta_{\vec{k}}(\vec{q}))
\theta(\Delta_{\vec{k}"}(\vec{q}))\sqrt{|\Delta_{\vec{k}}(\vec{q})
\Delta_{\vec{k}"}(\vec{q})|}(\alpha^*_{\vec{k}"\vec{k}'}+
\beta_{\vec{k}"\vec{k}'}).
\end{eqnarray}
   Solving these equations we find that in general there exists two kinds
of solutions: (i)particle-hole continuum, with $E_{\vec{k}}(\vec{q})=|
\vec{k}.\vec{q}|/m$, and (ii)collective modes, with energy $E_{0}(\vec{q})$
outside the particle-hole continuum satisfying the RPA eigenvalue equation
$1-v(q)\chi_0(\vec{q},E_0(\vec{q}))=0$. Most of the detailed mathematics can
be found in Ref\cite{b5}. We obtain finally,
\begin{mathletters}
\label{sol}
\begin{eqnarray}
\label{sol1}
\alpha_{\vec{k}\vec{k}'} & = & \delta_{\vec{k}\vec{k}'}+P{\theta(\Delta
_{\vec{k}}(\vec{q}))\theta(\Delta_{\vec{k}'}(\vec{q}))\sqrt{|
\Delta_{\vec{k}}(\vec{q})\Delta_{\vec{k}'}(\vec{q})|}v_{eff}(q,
|\vec{k}'.\vec{q}|/m)\over{L}^d({|\vec{k}'.\vec{q}|\over{m}}-{
|\vec{k}.\vec{q}|\over{m}})},   \\  \nonumber
\beta_{\vec{k}\vec{k}'} & = & {-\theta(\Delta_{\vec{k}}(\vec{q}))
\theta(\Delta_{\vec{k}'}(\vec{q}))\sqrt{|\Delta_{\vec{k}}(\vec{q})
\Delta_{\vec{k}'}(\vec{q})|}v_{eff}(q,-|\vec{k}'.\vec{q}|/m)\over
{L}^d({|\vec{k}'.\vec{q}|\over{m}}+{|\vec{k}.\vec{q}|\over{m}})}
\end{eqnarray}
for the particle-hole continuum spectrum $\vec{k}'$, where
$v_{eff}(q,\omega)=v(q)/(1-v(q)\chi_0(q,\omega))$ is the RPA
screened interaction, and
\begin{eqnarray}
\label{sol2}
\alpha_{\vec{k}0} & = & 
{1\over{L}^{d/2}}{\theta(\Delta_{\vec{k}}(\vec{q}))
\sqrt{|\Delta_{\vec{k}}(\vec{q})|}\over(E_0(\vec{q})-{|\vec{k}.\vec{q}|
\over{m}})\left[-{\partial\chi_0(q,\omega)\over\partial\omega}\right]
_{\omega=E_0(\vec{q})}^{1\over2}},  \\  \nonumber
\beta_{\vec{k}0} & = & 
-{1\over{L}^{d/2}}{\theta(\Delta_{\vec{k}}
(\vec{q}))\sqrt{|\Delta_{\vec{k}}(\vec{q})|}\over(E_0(\vec{q})+
{|\vec{k}.\vec{q}|\over{m}})\left[-{\partial\chi_0(q,\omega)
\over\partial\omega}\right]_{\omega=E_0(\vec{q})}^{1\over2}},
\end{eqnarray}
\end{mathletters}
for the collective mode $E_0(\vec{q})$. Notice that in the boson 
representation, the collective mode exists as an
non-perturbative effect due to interaction and cannot be obtained
from analytic continuation of the non-interacting boson modes.

   Next we examine the solutions of the bosonized Hamiltonian in the
$e^2\rightarrow\infty$ limit. First we consider the collective mode.
Using the result that $\chi_0(\vec{q},\omega)\rightarrow{n_oq^2\over{m}
\omega^2}$ in the limit $\omega>>k_Fq/m$, where $n_0$ is the fermion 
density\cite{mahan}, it is easy to see that
in the limit $e^2\rightarrow\infty$, the collective mode
frequencies are given by $E_0(q)=\omega_P$, where $\omega_P=
({4\pi{n}_0e^2\over{m}})^{1\over2}$ is the plasma frequency.
Notice that $\omega_P\rightarrow\infty$ as $e^2\rightarrow\infty$,
indicating that plasma oscillations are outside the physical spectrum 
in this limit. 

  Despite the vanishing of collective excitation in the physical
spectrum, the particle-hole excitation spectrum with excitation energies
$|\vec{k}.\vec{q}|/m$ survives in bosonization theory in
the limit $e^2\rightarrow\infty$.
In this limit $v_{eff}(q,|\vec{k}'.\vec{q}|/m)\rightarrow-1/
\chi_0(q,|\vec{k}'.\vec{q}|/m)$ and the coefficients $\alpha_{\vec{k}
\vec{k}'}$'s and $\beta_{\vec{k}\vec{k}'}$'s remain regular, indicating
that the particle-hole excitation spectrum is not qualitatively
modified by effect of confinement. It is instructive to
examine the charge fluctuations carried by the particle-hole excitations
by examining the commutator $[\rho(\vec{q}),\gamma^+_{\vec{k}}
(\vec{q}')]$. In particular, the commutator will be zero if
creating a particle-hole eigen-excitation does not introduce any
charge fluctuations in the system, which is what we expect in the
$e^2\rightarrow\infty$ limit. Using Eqs.\ (\ref{can}), \ (\ref{bgi})
and \ (\ref{sol1}), and the usual
boson commutation rules, it is straightforward to show that
\begin{equation}
\label{ccar}
[\rho(\vec{q}),\gamma^+_{\vec{k}}(-\vec{q}')]=\delta_{\vec{q}\vec{q}'}
\theta(\Delta_{\vec{k}}
(\vec{q}))\sqrt{|\Delta_{\vec{k}}(\vec{q})|}\times{1\over1-v(q)
\chi_0(\vec{q},|\vec{k}.\vec{q}|/m)},
\end{equation} 
and vanishes in the limit $e^2\rightarrow\infty$.

  Before ending this section let us examine our results obtained so
far from bosonization theory. Within the Gaussian approximation,
we find a RPA-like excitation spectrum with both collective modes
and particle-hole excitations. As the coupling constant $e^2$
increases, the energy of the collective mode rises continuously
to infinity whereas the particle-hole excitation spectrum remains
unaltered. The charge fluctuations carried by the particle-hole
excitations are projected out gradually as $e^2$ increases, resulting
in {\em chargeless} particle-hole excitations at $q<\Lambda$ in the confinement
limit $e^2\rightarrow\infty$. Notice that within the Gaussian approximation,
the confinement state analytically continues to the usual Fermi liquid
state and there is no phase transition in between. The theory thus
suggests that the confinement state of a gas of fermions is
a Fermi liquid state, with however {\em chargeless}-quasi-particles 
constituting the Fermi liquid.  It also suggests that
this is a rather unusual Fermi liquid state, since bare fermions in
the system carries charge $e$, and the quasi-particles must have
vanishing overlap with bare fermions if they carry zero charge.

\section{single-particle properties}
    In usual bosonization theory for one-dimensional systems, the
single particle properties of the system can be determined once a
rigorous representation of the single-particle operator in terms
of density operators $\rho_L(q)$ and $\rho_R(q)$ are obtained\cite{b1}. 
In higher dimensions, this procedure becomes inadequate for two
reasons: (1)The corresponding procedure requires that the bosonized
representation of single-particle operator $\psi_b(\vec{r})$
satisfies the commutation relations 
\[
[\psi_b(\vec{r}),\rho_{\vec{k}}(\vec{q})]=e^{-i(\vec{k}+\vec{q}/2).
\vec{r}}\int{d}^dr'e^{i(\vec{k}-\vec{q}/2).\vec{r}'}\psi_b(\vec{r}')
\]
for all possible momenta $\vec{k}$ and $\vec{q}$. We have not 
been able to find  a representation which satisfies this
criteria\cite{b2,b3,b4}, 
and even if we can find such can representation, the theory
is still approximate because in dimensions higher than one,
the boson representation using Wigner operators is not exact
and violates the Pauli exclusion principle, at
least in Gaussian approximation. (2)more importantly, unlike in one 
dimension where the elementary excitations are collective
density waves, we have seen in last section that in dimensions
higher than one the particle-hole excitation spectrum is fermi-liquid 
like, implying that fermionic quasi-particles exist in dimensions
higher than one. It is thus important to construct directly the 
quasi-particle operators in this case.

   To construct the quasi-particle operators we first consider
the equation of motion of
the bare fermion operator $\psi(\vec{r})={1\over{L}^d}
\sum_{\vec{k}}e^{-i\vec{k}.\vec{r}}f_{\vec{k}}$ at imaginary time,
\begin{eqnarray}
\label{emb}
{\partial\psi(\vec{r})\over\partial\tau} & = & {1\over2m}\nabla^2\psi
(\vec{r})-{1\over{L}^d}\sum_{\vec{q}}v(q)\rho(\vec{q})e^{i\vec{q}.\vec{r}}
\psi(\vec{r})  \\    \nonumber
& = & {1\over2m}\nabla^2\psi(\vec{r})-{1\over{L}^{d}}\sum_{\vec{q}}
\left[v(q)\rho_{ph}(\vec{q})+v(q)\rho_c(\vec{q})
\right]e^{i\vec{q}.\vec{r}}\psi(\vec{r}),
\end{eqnarray}
where
\begin{eqnarray}
\label{rho1}
v(q)\rho_{ph}(q) & = & \sum_{\vec{k}}\sqrt{|\Delta_{\vec{k}}(\vec{q})|}
\theta(\Delta_{\vec{k}}(\vec{q}))\left[v_{eff}(q,{-|\vec{k}.\vec{q}|
\over{m}})\gamma^+_{\vec{k}}(\vec{q})+v_{eff}(q,{|\vec{k}.\vec{q}|
\over{m}})\gamma_{-\vec{k}}(-\vec{q})\right],   \\  \nonumber
v(q)\rho_c(\vec{q}) & = & L^{d/2}\left(-{\partial\chi_0(q,\omega)
\over\partial\omega}\right)^{-{1\over2}}_{\omega=E_0(\vec{q})}
\left[\gamma_0^+(\vec{q})+\gamma_0(-\vec{q})\right],
\end{eqnarray}
where $v(q)\rho_{ph}(\vec{q})\sim\gamma^+_{\vec{k}}(\vec{q})+
\gamma_{-\vec{k}}(-\vec{q})$ describes the coupling of 
the particle-hole excitations to the fermion operator, and
$v(q)\rho_c(\vec{q})\sim\gamma_0^+(\vec{q})+\gamma_0(-\vec{q})$ 
describes coupling of the collective excitations (plasmons) to
the fermion operator. In the $e^2\rightarrow\infty$
limit, the interaction between fermions and particle-hole excitations
are regular and finite, and the strong confinement effect shows up only
in the interaction between fermions and collective mode excitations
(plasmons). The effect of plasmons on the one-particle
properties can be estimated perturbatively by evaluating the plasmon
contribution to fermion self-energy $\Sigma$ and renormalization
factor $z$ to second order using
Eqs.\ (\ref{emb}) and \ (\ref{rho1}). We find that $\Sigma\sim
\int^{\Lambda}_{L^{-1}}d^dqv(q)$, and $z\sim{\partial\Sigma\over
\partial\omega}\sim\Sigma/\omega_P$. Notice that both $\Sigma$ and
$z$ goes to infinity as $e^2\rightarrow\infty$. Furthermore, the
integrals carry also infrared divergence at dimensions $d\leq2$ for any
finite $e^2$. The divergence of $z$ at $d=2$ for finite $e^2$ was  
interpreted as signature of a marginal fermi liquid\cite{wen}.

  The divergence in single-particle self-energy indicates that 
the bare fermions is not a good starting
point for constructing quasi-particle operators. To find a better
starting point we first look
for a canonical transformation for the single-particle
operator which diagonalize the {\em interaction term} between fermions
and plasmons. The kinetic energy of fermions and interaction with
particle-hole excitations will be treated afterward.
We obtain\cite{sp}
\begin{mathletters}
\label{cano}
\begin{equation}
\label{qu1}
\psi_Q(\vec{r})=e^{\phi(\vec{r})}\psi(\vec{r}),
\end{equation}
where
\begin{equation}
\label{qu2}
\phi(\vec{r})={1\over{L}^{d/2}}\sum_{\vec{q}}
{e^{i\vec{q}.\vec{r}}\over{E}_0(\vec{q})
\left[-{\partial\chi_0(q,\omega)\over\partial\omega}\right]^{1\over2}}
(\gamma^+_0(\vec{q})-\gamma_0(-\vec{q})),
\end{equation}
\end{mathletters}
which represents a fermion operator "dressed" by plasmon modes.
It is straightforward to obtain the equation
of motion of the "dressed" fermion $\psi_Q$,
\begin{eqnarray}
\label{eqqs}
{\partial\psi_Q(\vec{r})\over\partial\tau} & = & {1\over2m}\left(
\nabla^2\psi_Q(\vec{r})-[\nabla^2\phi(\vec{r})-(\nabla\phi(\vec{r}))^2]
\psi_Q(\vec{r})-2\nabla\phi(\vec{r}).\nabla\psi_Q(\vec{r})\right)
\\   \nonumber
& & -{1\over{L}^d}\sum_{\vec{r}}v(q)\rho_{ph}(\vec{q})e^{i\vec{q}.\vec{r}}
\psi_Q(\vec{r}),
\end{eqnarray}
where we have used the results $[H,\gamma_0(\vec{q})]=-E_0(\vec{q})
\gamma_0(\vec{q})$ and $[\gamma_{\vec{k}}(\vec{q}),\gamma_0^+
(\vec{q})]=0$, etc. for $\vec{k}\neq0$ and the usual commutator 
between bare fermions and density operators to derive the above equation.
We have also neglected constant energy terms coming from
normal ordering of operators in Eq.\ (\ref{eqqs}). It is clear
from Eq.\ (\ref{eqqs}) that the direct coupling between fermions and
plasmons is eliminated in the equation of motion of $\psi_Q(\vec{r})$.
However, interaction between fermions and particle-hole excitations
remains in the equation of motion. Moreover, an indirect coupling to
plasmon is also generated from the "dressed" fermion kinetic energy
term as is in the similar "small polaron" problem\cite{sp}. It is easy 
to see by direct power counting of $e^2$ in the $\phi(\vec{r})$
field that the coupling of the "dressed" fermion to plasmons 
through kinetic energy term is much
weaker than the original fermion-plasmon coupling. In particular,
we find that the self-energy correction of "dressed" fermions
from $\phi(\vec{r})$ fields remains finite in the limit
$e^2\rightarrow\infty$. It is also straightforward to show that the
infra-red divergence in fermion self-energy
at two dimensions is removed for the dressed fermions because of the
much weaker coupling to plasmons.

   These results suggest that the dressed fermion operators
$\psi_Q(\vec{r})$'s constitute a valid starting point to 
construct chargeless quasi-particles in the $e^2\rightarrow\infty$
limit. The other interaction effects can be treated perturbatively.
To show that this is indeed the case we first consider the commutation
relation between charge and dressed fermion operators.
It is straightforward to show that
\begin{equation}
\label{cless}
[\rho(\vec{q}),\psi_Q(\vec{r})]=\left({2\chi_0(q,E_0(\vec{q}))\over
E_0(\vec{q})\left[-{\partial\chi_0(q,\omega)\over\partial\omega}
\right]_{\omega=E_0(\vec{q})}}-1\right)e^{-i\vec{q}.\vec{r}}
\psi_Q(\vec{r}),
\end{equation}
which vanishes in the limit $e^2\rightarrow\infty$, when $E_0(\vec{q})
\rightarrow\omega_P\rightarrow\infty$, indicating that the dressed
single particle operators $\psi_Q(\vec{r})$'s are indeed 'chargeless'
in the $e^2\rightarrow\infty$ limit. To show that $\psi_Q(\vec{r})$
and $\psi_Q(\vec{r}')$ defines independent quasi-particles when
$\vec{r}\neq\vec{r}'$ we check the commutation relation
between the dressed fermion operators themselves. We obtain
\[
[\psi_Q(\vec{r}),\psi_Q^+(\vec{r}')]\sim{1\over{n}_0(\pi|\vec{r}-
\vec{r}'|)^{d-1}}\hat{O}(\vec{r},\vec{r}'),   \]
in the limits $e^2\rightarrow\infty$ and $|\vec{r}-\vec{r}'|\rightarrow
\infty$, where 
\[
\hat{O}=({\vec{r}-\vec{r}'\over|\vec{r}-\vec{r}'|}).\left[
\psi^+(\vec{r}')e^{\phi(\vec{r})}(\nabla\psi(\vec{r}))e^{\phi^+(\vec{r}')}
-(\nabla\psi^+(\vec{r}'))e^{\phi(\vec{r})}\psi(\vec{r})e^{\phi^+
(\vec{r}')}\right].  \] 
The vanishing of the
commutator between different $\psi_Q$ operators separated by 
large distances at dimensions larger than one indicates
that they can be used to construct independent quasi-particle operators 
when describing the dynamics of these systems at long distance.
Notice that in one dimension, such a construction is not possible
because of the long-rangeness of the commutation relation. In fact,
the only fermionic operators which commute with density operators
$\rho(\vec{q})$'s are the ladder operators\cite{b1} which raise or lower
the number of particles in the system by one. There are only two 
independent ladder operators in the system, which change the number of 
left-going and right-going fermions and cannot be used to construct
local quasi-particle excitations.

Finally, it is also easy to show that
\[
<\psi^+_Q(\vec{r})\psi_Q(\vec{r})>=<\psi^+(\vec{r})\psi(\vec{r})>.  \]
The fact that the densities of bare fermions and dressed fermions are the
same implys that the fermi surface volume of the dressed fermions is
exactly the same as that of the bare fermions. Assuming that the dressed
fermions form a free fermi gas, we also obtain
\[
<\psi^+_Q(\vec{r})\psi(\vec{r})>\sim{e}^{-({\pi{m}e^2\over{n}_0})^{1\over2}
(\int^{\Lambda}_{L^{-1}}d^dq{1\over{q}^2})}<\psi^+(\vec{r})\psi(\vec{r})>, 
 \]
to leading order in $e^2$, which vanishes in the 
$e\rightarrow\infty$ limit, indicating that the bare fermions
and dressed fermions have zero wavefunction overlap, as is expected
on physical ground. A similar calculation shows that the bare-fermion
occupation number $n(\vec{k})$ has no discontinuity across fermi
surface in the $e^2\rightarrow\infty$ limit, in agreement with
marginal fermi liquid picture.

\section{summary}
   Using a bosonization approximation we studied in this paper
a gas of fermions interacting {\em via} scalar potential $v(q)=
4\pi{e}^2/q^2$ for $q<\Lambda<<k_F$. In particular we consider the
$e^2\rightarrow\infty$ limit where the potential becomes confining. 
We note that because of the low momentum cutoff $\Lambda<<k_F$ in
our model, a crystal state is not expected to be formed and
a new treatment of the problem is necessary. Within a Gaussian
approximation we find that the particle-hole excitation spectrum
of the system is always fermi-liquid like, with the charge carried
by the particle-hole excitation vanishing continuously in the
$e^2\rightarrow\infty$ limit. Based on this result chargeless 
fermionic operators are construct which, we believe can be used
as the starting point for constructing quasi-particles in the system.
The system can be considered as a 'marginal' fermi liquid where
the chargeless quasi-particles have vanishing wavefunction
overlap with bare fermions in the system. 

  A major assumption we have made in our theory is that the
Gaussian approximation for the Wigner bosons describes at least
qualitatively correct the particle-hole excitation spectrum of
the system. The Gaussian approximation is believed to be valid
as long as the interaction potential is restriected to region of
small momentum transfer $q<\Lambda<<k_F$\cite{b2,b3,b4}. Notice that
the approximation can be improved systematically
{\em via} a cummulant expansion\cite{b4}. We find that to lowest
order of the cummulant expansion there is no extra divergence
introduced into the particle-hole excitation spectrum\cite{ng}.
Notice also that we have considered here only the effect of
longitudinal confining gauge fields in gas of fermions. Naively
we expect that similar physics will be found in the presence of
transverse gauge field. In the $e^2\rightarrow\infty$ limit,
the chargeless quasi-particles we have constructed will decouple
from the gauge fields and form a (marginal) fermi liquid. However,
it is easy to show that the $e^{i\phi(\vec{r})}$ operator we have
constructed is not able to "screen out" the transverse gauge field
because it is constructed from density fluctuations which couples to
longitudinal gauge field only. The transverse gauge fields requires 
rather different treatment and we shall discuss it in a separate paper.

  The author thanks Prof. Zhao-bin Su for many helpful questions and
comments. This work is supported by HKUGC through RGC grant HKUST6124/98P.


\begin{references}
\bibitem{t1} N. Nagaosa and P.A. Lee, \prl {\bf 60}, 2450(1990); X.G.
  Wen and P.A. Lee, \prl {\bf 76}, 503(1996).
\bibitem{t2} B.I. Halperin, P.A. Lee and N. Read, \prb {\bf 47}, 7312
 (1993); Y.B. Kim, P.A. Lee and X.G. Wen, \prb {\bf 52}, 17275 (1995).
\bibitem{hal}, see A. Stern, B.I. Halperin, F. von Oppen and S.H. Simon,
   cond-matt/9812135 for a review.
\bibitem{b0} S. Tomonaga, Prog. Theor. Phys. {\bf 5}, 544(1950);
 D.C. Mattis and E.H. Lieb, J. Math. Phys. {\bf 6}, 304(1965).
\bibitem{b1} see for example, F.D.M. Haldane, J. Phys. C {\bf 14},
 2585(1981).
\bibitem{b2} F.D.M. Haldane, Helv. Phys. Acta {\bf 65}, 152(1992).
\bibitem{b3} A. Houghton and J.B. Marston, \prb {\bf 48}, 7790(1993);
  A.H. Castro Neto and E. Fradkin, \prl {\bf 72}, 1393(1994).
\bibitem{b4} P. Kopietz, J. Hermisson and K. Sch$\ddot{o}$nhammer,
 \prb {\bf 52}, 10877(1995).
\bibitem{gauge} D.V. Khveshchenko and P.C.E. Stamp, \prl {\bf 71},
 2118 (1993); see also B.L. Altshuler, L.B. Ioffe and A.J. Millis,
 \prb {\bf 50}, 14048 (1994).
\bibitem{b5} A.H. Castro Neto and E. Fradkin, \prb {\bf 49}, 10877(1994).
\bibitem{mahan} see for example, G.D. Mahan, in {\em Many-Particle Physics},
  (Plenum Press, New York and London (1990)).
\bibitem{wen} P.A. Bare and X.G. Wen, \prb {\bf 48}, 8636(1993).
\bibitem{sp} notice that a similar result is also obtained in the case of
   small polaron problem; see ref.[8] for example.
\bibitem{ng} T.K. Ng, unpublished.
\end{references}
\end{document}